# BITTORRENT NETWORK TRAFFIC FORECASTING WITH ARMA

Poo KuanHoong, Ian K.T. Tan and CheeYikKeong

Faculty of Computing and Informatics, Multimedia University,
Persiaran Multimedia, Cyberjaya, 63100, Selangor D.E., Malaysia
khpoo@mmu.edu.my, ian@mmu.edu.my, cyknmk@gmail.com

## ABSTRACT

*In recent years, there are some major changes in the way content is being distributed over the network. The content distribution techniques have recently started to embrace peer-to-peer (P2P) systems as an alternative to the traditional client-server architecture. P2P systemsthat are based on the BitTorrent protocol uses end-users' resources to provide a cost effective distribution of bandwidth intensive content to thousands of users. The BitTorrent protocol system offers a scalable mechanism for distributing a large volume of data to a set of peers over the Internet. With the growing demand for file sharing and content distribution, BitTorrent has become one of the most popular Internet applications and contributes to a signification fraction of the Internet traffic. With the wide usage of the BitTorrent protocol system, it has basically solved one of the major problems where data can be quickly transferred to a group of interested parties. The strength of the BitTorrent protocol lies in efficient bandwidth utilization for the downloading and uploading processes. However, the usage of BitTorrent protocol also causes latency for other applications in terms of network bandwidth which in turn has caused concerns for the Internet Service Providers, who strives for quality of service for all their customers. In this paper, we study the network traffic patterns of theBitTorrent network traffic and investigate its behavior by usingthe time series ARMA model. Our experimental results show that BitTorrent network traffic can be modeled and forecasted by using ARMA models. We compared and evaluated the forecasted network traffic with the real traffic patterns. This modeling can be utilized by the Internet Service Providers to manage their network bandwidth and also detect any abnormality in their network.*
.

## 1. INTRODUCTION

BitTorrent [1] is one of the most popular file sharing protocols that is widely used for file sharing and content distribution. BitTorrent uses peer-to-peer (P2P) sharing protocol that has shown tremendous success in distributing large files efficiently over wide area networks. Its efficiency and speed for file distribution has resulted inBitTorrent clients being created for all common operating systems using a variety of programming languages [2]. For BitTorrent file sharing, research study shows that the serving capacity of the BitTorrent protocol grows as number of nodes increase and it leverages on the bandwidth contribution by all the connected peers. In order to ensure high availability, BitTorrent downloads in a random or in a "rarest-first" approach where it sends multiple requests to many available existing nodes and looks for the opportunity to download chunks (blocks) of files that are scarce or rare to others. However, this approach hasa drawback in terms of network traffic where this will cause network traffic congestion. The bandwidth utilization by the BitTorrent traffic is significant especially when a peer has to download files in chunks of files from other peers and at the same time serves to contribute its upload bandwidth by uploading chunks of files to other connected peers in a P2P environment. As a result, this generates high volume of overall network traffic and will cause packet latency or packet loss. Even a network user, who is not using the BitTorrent file sharing protocol, may experience packet latency or packet loss if it is connected to the same local area network where one or more of the network users are transmitting data via a BitTorrent client application.

For Internet Service Providers (ISP) and telecommunication services providers, the tremendous growth of P2P traffic, especially in the BitTorrent usage, has caused network management issues where non-P2P users' traffic has been affected unfairly [3][4][5]. Not only do these affect the individual ISPs but also traffic between ISPs [5] and the involvement of regulatory bodies to ensure fairness to the ISPs' customers [3]. Several past researches have been done to address this issue by localizing the P2P traffic [6][7][8] through the reduction of traffic between ISPs. Bindalet. al. [6] concluded that if a peer chooses the majority of their neighbors, but not all, from within the same ISP, the downloading performance would be significantly improved. Building on this, each ISP can analyze the BitTorrent traffic pattern within their own networks to assist in ensuring a level of quality of service to their customers through careful network planning in terms of smart throttling for appropriate users. In understanding the BitTorrentprotocol, it uses a central location to manage users' downloads. The central location is known as a tracker. Clients connect to the tracker when a downloading process if required, which is done through the launching of a torrent file. The tracker keeps track of all the nodes that have the file (both partially and completely) and connects nodes to each other for downloading and uploading. Generally, there are two types of nodes in a P2P system, namely seeds and leechers. Seeds are nodes that have the complete block of files and ready to upload to other connected nodes while leechers are nodes that are currently downloading.

In this paper, we focus on determining whether it is possible to model the characteristic and network traffic pattern of seed nodes by using time series model as well as whether it is possible to forecast BitTorrent network traffic in order for ISPs to improve overall network traffic across their network and minimize inter-ISP network packets. In order to forecast the network traffic, we apply the Autoregressive Moving Average (ARMA) time series model. In terms of network traffic samples, we collect several sets of BitTorrent network traffic and apply the ARMA model on the log return values of the sampling. The forecast accuracy is determined by two factors which are ARMA $(p, q)$ where $p$ and $q$ are the orders, and Mean Square Error (MSE) on the forecasted results. In order to model the BitTorrent network traffic, we propose our own prediction model based on the behavior of BitTorrent network traffic captured over a short period. In this paper, the data are processed in terms of number of packets while the log return values show the variation of packets throughput over a unit of time (second).

The main contributions of this paper are as follows:

- Prediction models for both seasonal and cyclical BitTorrent network traffic patterns are proposed.
- Based on the proposed prediction models, BitTorrent network traffic were forecastedusing the captured network traffic.
- The results were analysed and shown that our proposed prediction models can effectively be applied for BitTorrent network forecasting.

The remainder of this paper is organized as follows: Section II discusses about ARMA and others related work with the ARMA model. Our implementation work is explained in section III, we analyze and discuss our experiment results in section IV and we conclude our findings in section V.

## 2. AUTOREGRESSIVE MOVING AVERAGE (ARMA)

The Autoregressive Moving Average (ARMA) is a time series model where the AutoRegressive (AR) component is the sum of past observations with the addition of a constant and white noise error value combined with the Moving Average (MA) component, which is the sum of the past white noise error terms in addition with the expected value of the time series and the white noise error value. With both the AR and MA components, the period for the past observations is defined by the order value and this is generally written as ARMA $(p, q)$ where $p$ is the order for the AR component and $q$ is the order for the MA component.

The AR is described by the $p$ sequence:

$$X_t = \sum_{i=1}^{p} \theta_i X_{t-1} + \varepsilon_t \qquad (1)$$

where $t \in \{0,1,2, \dots \}, \theta_1 \dots \theta_p$ are the parameters of the model that need to be determined, and we use $\varepsilon_t$ to represent the constant and white noise error value.

MA is described by $q$ sequence:

$$X_t = \mu + \sum_{i=1}^{q} \varphi_i \varepsilon_{t-i} + \varepsilon_t \qquad (2)$$

where $t \in \{0,1,2, \dots \}, \varphi_1 \dots \varphi_q$ are the parameters of the model that need to be determined, $\mu$ is the expected value of $X_t$ and $\varepsilon_t$ represents the white noise error value.

In the linear combination of both (1) and (2), assuming that the expected value $\mu$ is 0 (no error), it will result in;

$$X_t = \sum_{i=1}^{p} \theta_i X_{t-1} + \sum_{i=1}^{q} \varphi_i \varepsilon_{t-i} + \varepsilon_t \qquad (3)$$

where $t \in \{0,1,2, \dots \}, \theta_1 \dots \theta_p$ are the parameters for the AR model, $\varphi_1 \dots \varphi_q$ are the parameters for the MA model, $\varepsilon_t$ represents white noise error value, $p$ is the AR term and $q$ is the MA term.

ARMA $(p, q)$ models can be used to describe a wide variety of behaviors for stationary time series [9]. Forecasts generated by an ARMA $(p, q)$ model will depend on historical data as well as the $p$ and $q$ orders [10][11].

ARMA was initially applied for financial data forecasting [12] before it was adapted for network traffic modelling. Further work by P. Branch et. al. [13]showed that ARMA model is able to capture the behavior of network gaming traffic. ARMA model was applied on first person shooting game traffic to predict packet sizes distribution in multiplayer system where game state information is shared among players in real time. By using time series prediction model, they demonstrated that it is possible to improve the overall game state updates and minimize the probability of latency [13]. In other research work done by M. Papadopouliet. al. [14], they employed time series model to solve network traffic congestion issue ona wireless access point. Based on their time series model, it shows a positive result in fixing access point congestion period in a wireless environment. While, in the study by S. Jung et. al., they developed network measurement of China Educational Research NETwork (CERNET) northwest center and they presented and evaluated their results of ARMA and Autoregressive Integrated Moving Average (ARIMA) for long term prediction [15]. However, based on their findings, they conclude that ARMA model is best applied for short term prediction rather than long term prediction. Besides ARMA model, KiarashMizanianet. al. [16] has applied Ornstein Uhlenbeck (OU) process to model the bandwidth in P2P networks.They conducted studies on the auto covariance structure of bandwidth and the statistical parameters such as mean, variance and auto covariance are obtained.

## 2.1 Network Traffic Forecasting

Network traffic forecasting in computer and communication networks has become a very complex process. In order to forecast network traffic more accurately, it is dependent on two main parameters: (1) the accessibility of previous network traffic history data and (2) how far into the future a traffic rate process can be predicted for a given error constraint. Krithikaivasan et.al.[17] developed a seasonal Auto Regressive Conditional Heteroskedasticity (ARCH) based model with the innovation process (disturbances) generalized to the class of heavy-tailed distributions for non-stationary periodical measured traffic data. Their approach provides a robust dynamic bandwidth provisioning framework for real-world periodically measured non-stationary traffic. By incorporating neural networks and time series methods, P. Cortez et. al. [18] has shown that it is possible to forecast accurately the amount of traffic in TCP/IP based networks. They combined neural networks with two important adapted time series methods ARIMA and Holt-Winters. From their experimental results, the neural ensemble achieved the best resultsfor 5 minutes and hourly data, while the Holt-Winters is the best option for the daily forecasts.Other researchers have shown that network traffic forecasting can be further enhanced with time series models and artificial intelligence [19-22].Considering all of the factors, most research works proposed forecasting model that can utilized in order to forecasting network traffic for various network environments. Evaluations were conducted and results showed that traffic prediction can be utilized in network control areas.

## 3. IMPLEMENTATION

In order to capture BitTorrent network traffic, a P2P environment consists of a tracker, seed nodes and leecher nodes. The tracker sends packets across the network to keep track of all the nodes that are currently in the system. All nodes will exchange updates with other nodes in a timely manner.  As time progresses, the overall network traffic increases when a seed node receives request from other connected nodes and it will start to upload chunks of files to its connected neighbors. In this work, we collected network traffic from a computer (PC) which is connected to a campus wide network. In order to minimize the traffic overheads with other types of network traffic, we collected the sampling network traffic by seeding 10 torrents in a machine and capture their network packets for the period of 3 hours using Wireshark [23], a popular open source network protocol analyzer. This process is repeated for six days in order to obtain 6 sets of sampling data. BitTorrent commonly uses Transmission Control Protocol (TCP) and User Diagram Protocol (UDP) over Internet Protocol (IP) for transport. Therefore, the network traffic that we captured is filtered for only TCP and UDP packet-oriented protocols.

As for the prediction scenario, we applied a differencing technique on the sampling data in order to obtain a stationary time series. We apply the log return transformation on our sampling data to eliminate exponential data as well as to establish a stationary state. We developed a Simple Network Prediction Studio application (SNPS) which has the built-in features of log differencing and the ARMA model. This simplified application is able to predict log return values that we need to evaluate theBitTorrent traffic pattern. The tools that we use are described in subsection 3.1 and 3.2. Our implementation procedure is shown in Figure 1.

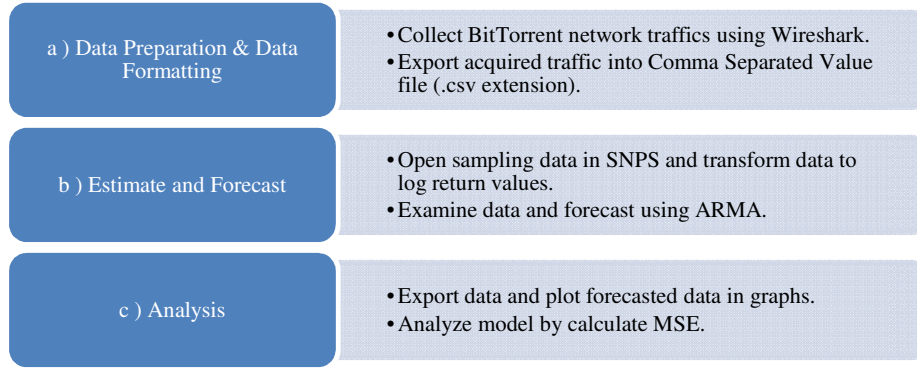

**FIGURE 1:** Implementation procedure of applying ARMA model on BitTorrent network traffic.

## 3.1 Wireshark

Wireshark [23] is an open source network protocol analyzer that runs on a variety of operating systems. Wireshark is highly customizable, allows for the filtering of network traffic according to the required specifications and the ability to export the output into a variety of formats. For our experiments, we configured Wireshark on our machine to capture network packets and export the captured data file every hour to be saved in order to maintain data consistency due to our memory buffer limitation of our machine. The data collected is exported in Comma Separated Value (.csv) format which consists of sequence number, time, source, destination, protocol and related information fields. In our work, we run Wireshark on Ubuntu, a Linux based operating system, where Wireshark is configured to capture based on a specific network interface and filtered by the string tcp or udp. It is observed that Wireshark consumes large memory buffer as time progresses and it is limited by the machine's memory capacity. Thus for long period of network packet capture, memory buffer is required to be cleared by exporting the collected data to multiple .csv files.

## 3.2 Simple Network Prediction Studio (SNPS)

The research work of Tan et. al. [24] uses the CRONOS [25] ARMA model application which is available as an Open Source Time Series Analysis project. We developed an application called Simple Network Prediction Studio (SNPS) using the base model provided by CRONOS. The SNPS application is designed to simplify the forecasting job where it can be executed more efficient via a simple interface.

The SNPS consists of four main features:

a) Transform Comma Separated Value (.csv) file into packets throughput/number of packets over time (seconds).
b) Log return transformation.
c) Forecast network traffic with different sampling (window) size, varying step size and different ARMA ($p$, $q$) orders.
d) Graph representation of forecasted results.

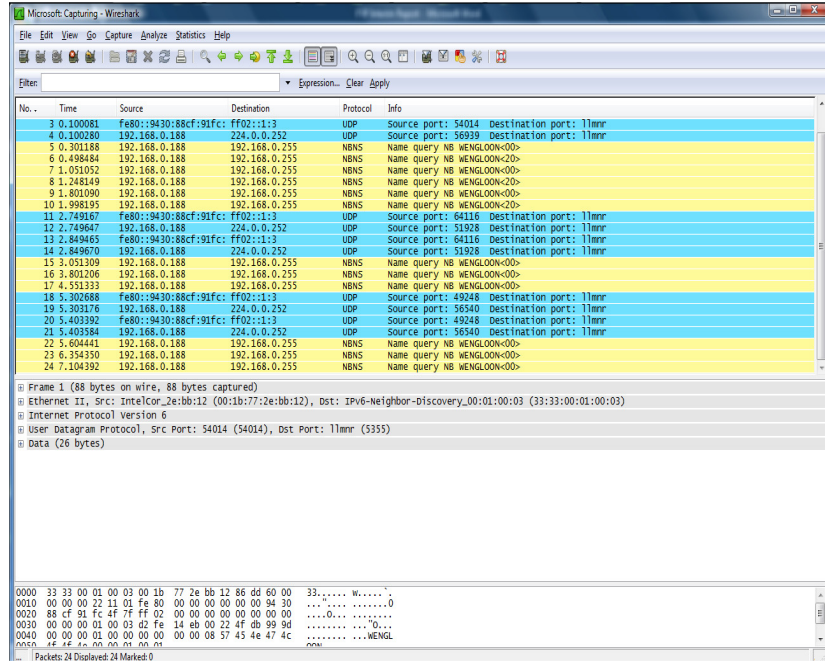

**FIGURE 2:** Wireshark Application Interface.

There are 3 important parameters in network traffic prediction. First, the "sampling" (or window) size is referred as the size of the historical data used in the model for forecast. It varies according to user's input and the size range within the maximum size of the sampling. While "step size" refers as the size in time (seconds) unit which tabulates the number of packets per step. Lastly, ARMA ($p$,$q$) refers to the $p$ and $q$ order that varies in forecasting the different categories of the time series.

Although SNPS supports the plotting of graphs automatically from the output, we exported our data and generate the graphs using Microsoft Excel as it will also allow easier computation of the Mean Square Error (MSE).

### 3.3 Model Prediction Scheme

In order to achieve high accuracy for network traffic forecasting, an accurate traffic prediction model is needed to avoid any overestimate or underestimate in the network performance. The proposed model should have the ability to capture the prominent traffic characteristics, e.g. short and long range dependency, and self-similarity. For our work, we designed an ARMA model scheme to study the behavior of BitTorrent traffic. ARMA model can be used to predict one step size ahead values of the time series, which can be extended to $k$-step size ahead. In this paper, we presented 5 step sizes ahead prediction for BitTorrent network traffic with 6 sets of sampling data. Our estimation target is the prediction of 15 step sizes ahead of values by using 5 step sizes ahead prediction technique with step size of 30 seconds as implemented by Tan et. al. [24]. Window size that serves as past historical data size is used in the forecast model, where the window size is varied randomly in order to obtain better forecast performance.

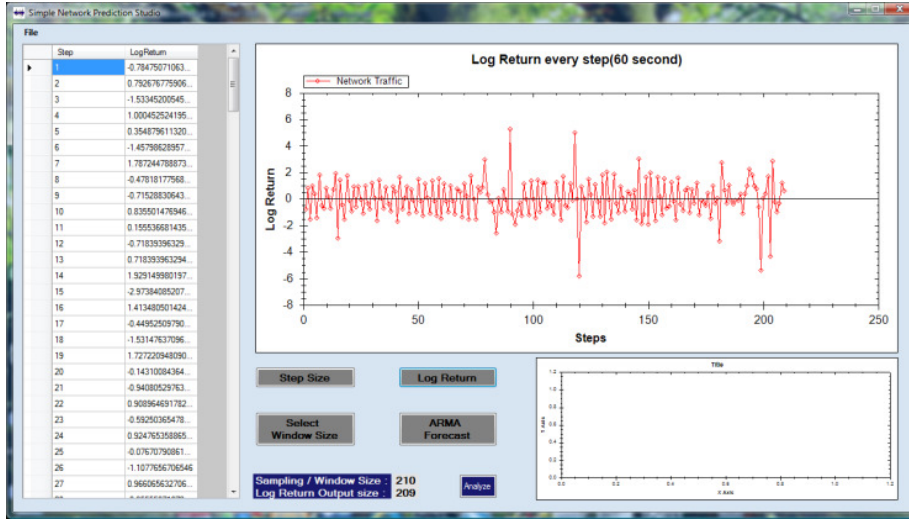

**FIGURE 3:** SNPS application interface.

Based on the time series pattern study by Qiuet. al. [9], we can categorize network traffic patterns into two; cyclical and seasonal patterns. The cyclical pattern is a time series that contains wavelike fluctuations around the trend component. While the seasonal pattern consists of repetitive patterns. We categorized BitTorrent traffic after the traffic data has been transformed using log return.

We used ARMA (1, 0) as suggested by the research done by Jung et. al. [15]. For reason of accuracy, ARMA (1, 0) is not a preferable order after several performanc testing on BitTorrent traffic. We amplify our $p$ order to 3 to evaluate its performance to study the behavior of the historical data. Work done by Wang et. al. [26] shows that ARMA (2, 1) build seasonal forecasting well and match the pattern approximately. We draw a scheme to illustrate our procedure on determining ARMA $p$ and $q$ order.

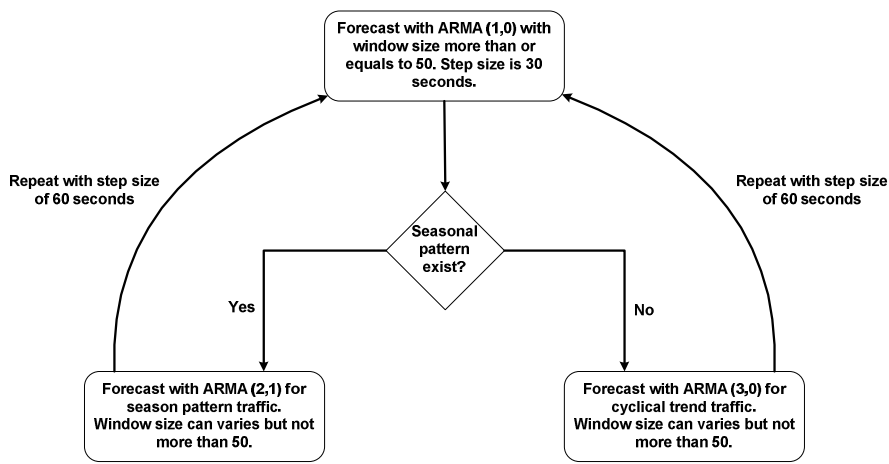

**FIGURE 4:** ARMA prediction scheme.

In order to evaluate the accuracy of our forecasting, The mean square error (MSE) of the log return is used to compare the precision of our proposed model.

$$MSE = \frac{1}{n} \sum (value_a(i) - value_p(i))^2$$

where $value_a$ denotes the $i^{th}$ of actual log return and $value_p$ denotes the $i^{th}$ of predicted log return.

# 4. RESULTS AND DISCUSSION

For our experiment data collections, we collected 6 sets of data over 6 days and applied ARMA (3, 0) with ARMA (2, 1) to the transformed network values. Based on the collected data, graphs were plotted for the 6 datasets. As shown in the graphs, we classified the graphs into seasonal and non-seasonal patterns as depicted in Table 1.

| Figure | Dataset | Seasonal Pattern |
|---|---|---|
| Figure 5a | A | No |
| Figure 5b | B | No |
| Figure 5c | C | No |
| Figure 5d | D | No |
| Figure 5e | E | Yes |
| Figure 5f | F | Yes |

**TABLE 1:** Tabulation of Network Forecasting Results.

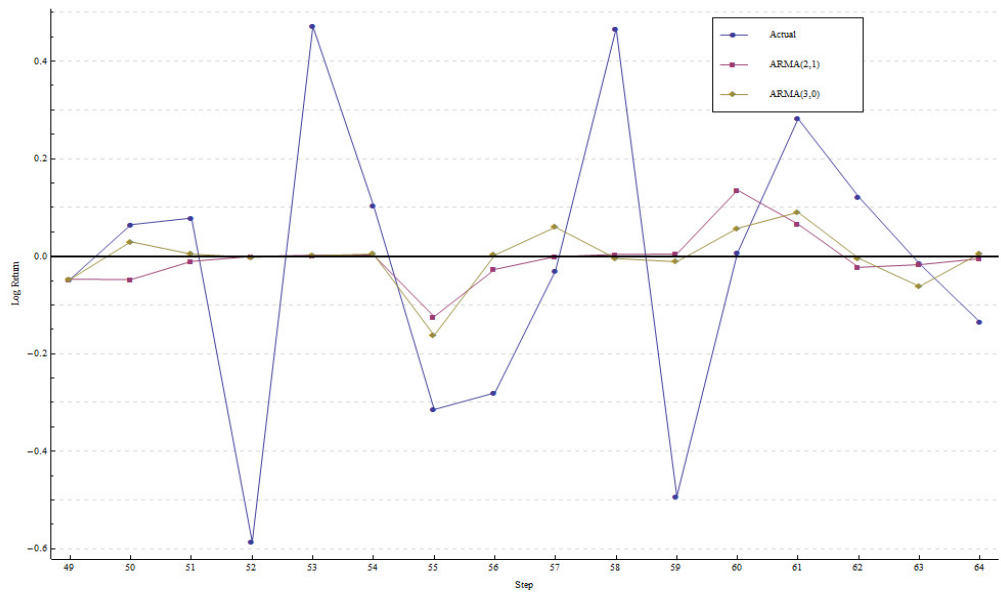

**FIGURE 5a:** Dataset A.

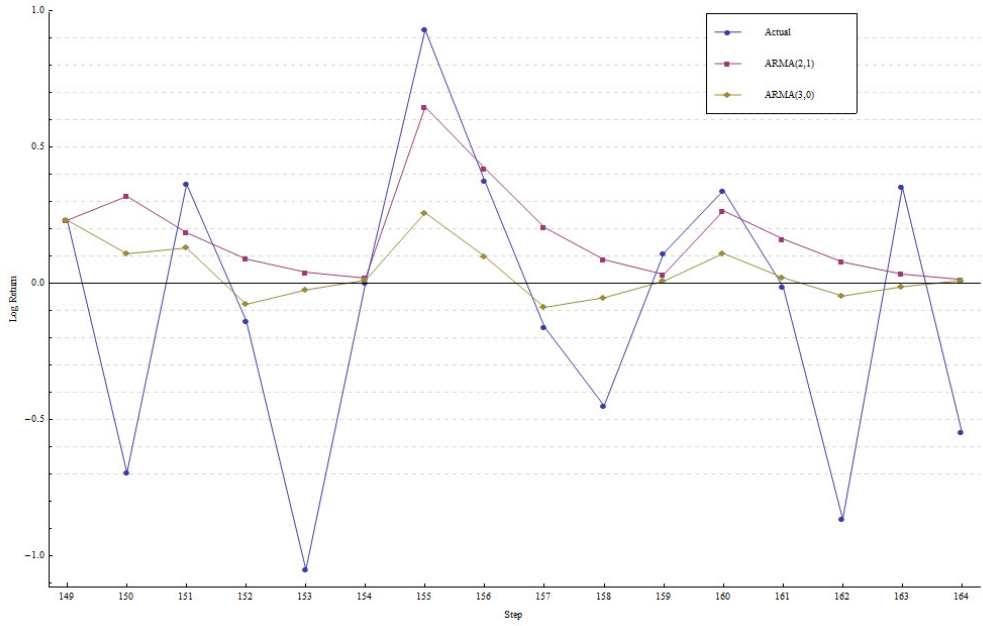

**FIGURE 5b:** Dataset B.

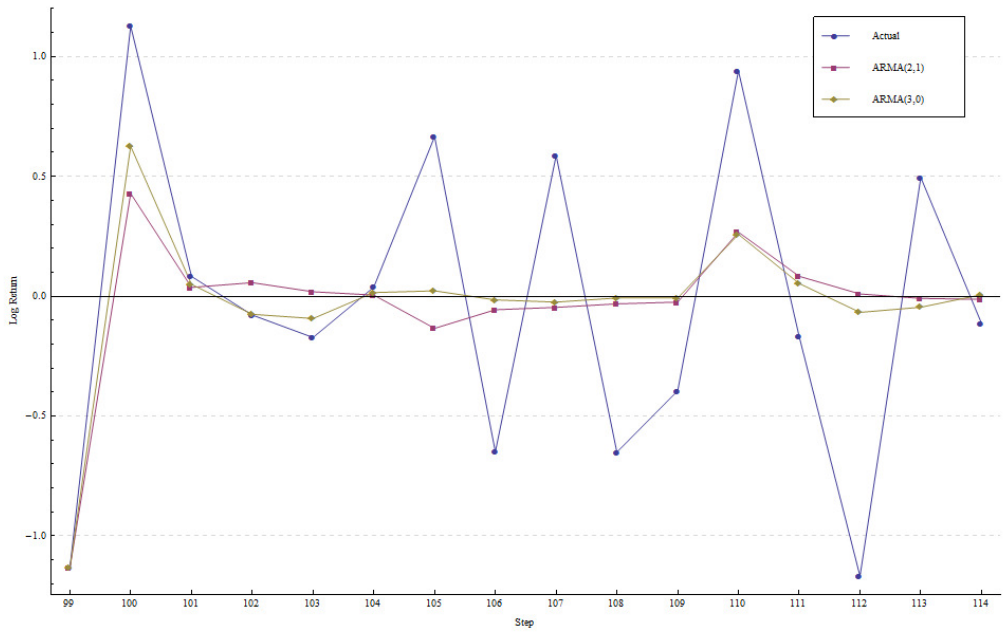

**FIGURE 5c:** Dataset C.

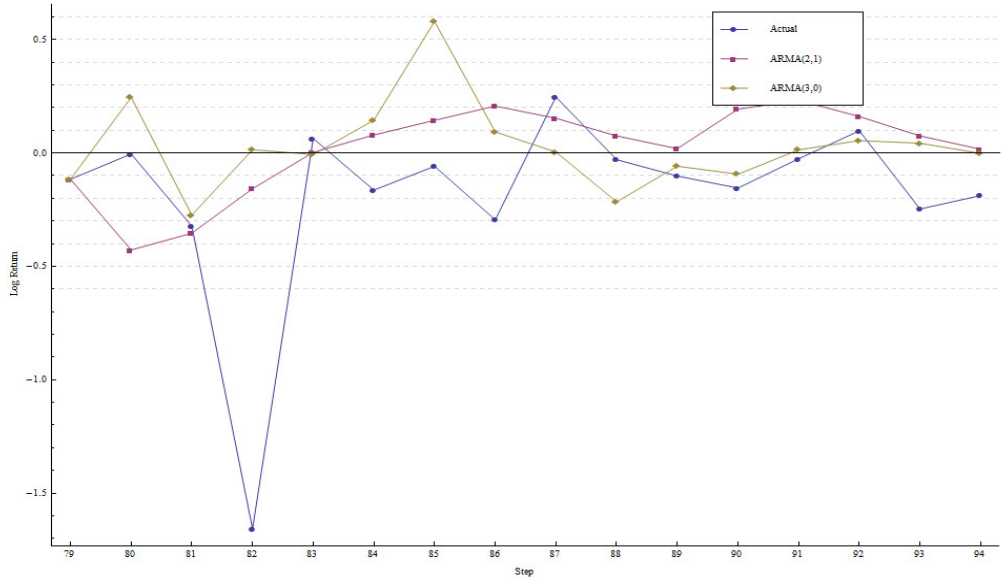

**FIGURE 5d:** Dataset D.

Datasets A to D (as depicted in Figures 5a to 5d) are classified as cyclical time series pattern where the traffic volume rises and falls does not follow any periodic patterns. These are also not influenced by the BitTorrent user requests nor the number of nodes in the BitTorrent system.

| Figure | MSE (ARMA(2,1)) | MSE (ARMA(3,0)) |
|--------|------------------|------------------|
| 5a     | 0.083902         | 0.081404         |
| 5b     | 0.277203         | 0.241949         |
| 5c     | 0.308422         | 0.269801         |
| 5d     | 0.210333         | 0.249064         |

**TABLE 2:** Mean Square Error (MSE) for Datasets A to D.

We computed the Mean Square Error (MSE) for both ARMA (2, 1) and ARMA (3, 0) and in these cyclical traffic patterns and noted that the average MSE value for ARMA (3, 0) is lower as compared to ARMA (2,1). We can conclude that ARMA (3,0) outperforms ARMA (2, 1). The ARMA (3, 0) is better suited for representing the growth and decline of the traffic volume of the cyclical BitTorrent network traffic.

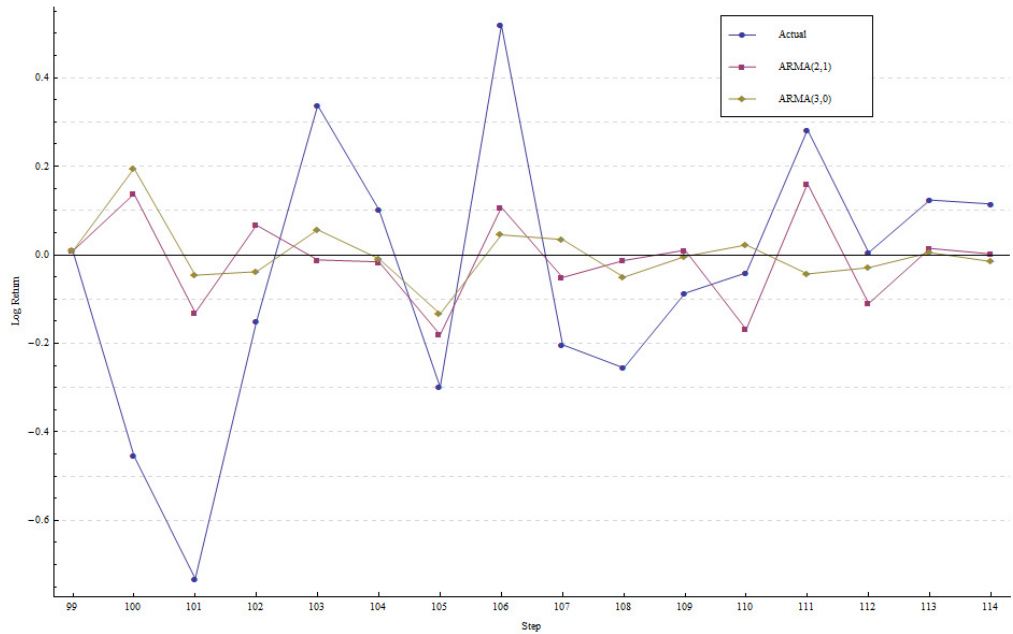

**FIGURE 5e:** Dataset E.

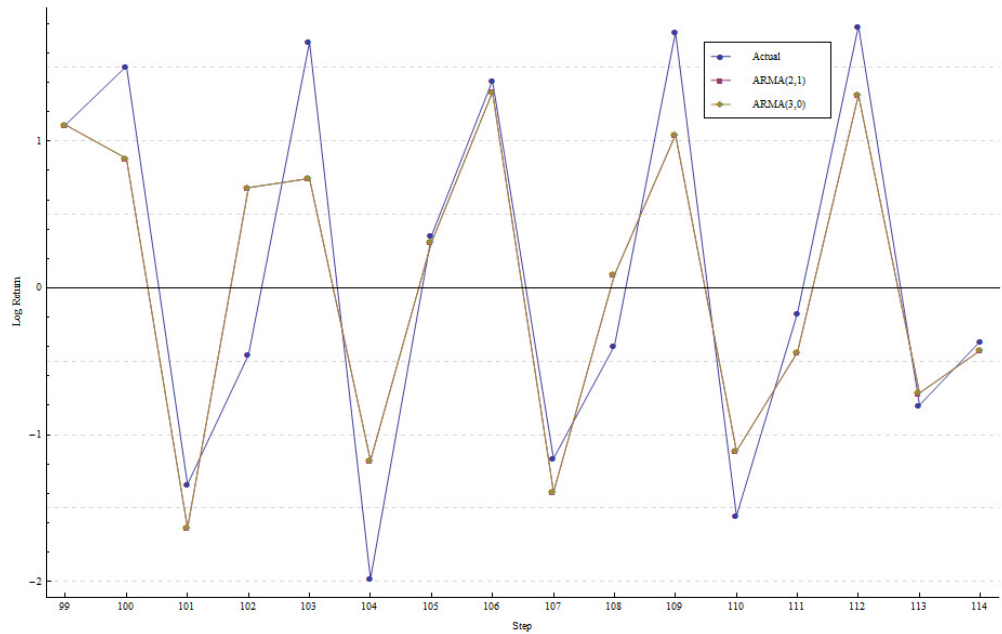

**FIGURE 5f:** Dataset F.

From the plotted graphs of figure 5e and 5f, it is observed that the data has seasonal patterns. For forecasting seasonal patterns, the Moving Average (MA) *q* order has the strongest influence due to using the same weights (of the order) for each observation (step) without any references to the peak and valley of the graphs [26]. This applies to BitTorrentnetwork traffic as well where the network received similar request, nodes updates and constant uploading to leechersthrough the network. Thisbehavior explains the reason seasonal BitTorrent network traffic happens and ARMA (2, 1) is observed to be able to predict accurately.

| Figure | MSE (ARMA(2,1)) | MSE (ARMA(3,0)) |
|--------|-----------------|-----------------|
| 5e | 0.08256 | 0.099679 |
| 5f | 0.304797 | 0.304815 |

**TABLE 3:** Mean Square Error (MSE) in comparison.

Table 3 lists the Mean Square Error (MSE) for ARMA (2, 1) and ARMA (3, 0) for Datasets E and F. The results indicate the correctness of our scheme that there are differences between ARMA *p* and *q* order for different time series pattern. The mean square error becomes larger when we use ARMA (2, 1) to forecast a cyclical time series pattern.

Our experimental results show that large variability of cyclical time series pattern makes it difficult to predict as ARMA prediction works within a certain limit. Another observation is that the prediction of BitTorrent traffic achieve better with seasonal pattern compared to cyclical pattern, mostly due to the seasonal trend contains in time series can be captured and evaluated using ARMA model.

## 5. CONCLUSION

In this paper, we presented evidence which suggests that the BitTorrent network traffic can be understood and hence can be short term forecasted using the ARMA model. The data collected has also shown that the traffic generated by the BitTorrent seeds can either be seasonal or cyclical pattern. By using the ARMA time series prediction, other types of network activities can be scheduled for the time period when the system predicts a low BitTorrent network traffic. This will improve the users' application bandwidth usage with lesser occurrences of network congestion. This in turn implies that the users achieve better quality of service from their ISPs.

In summary, our research findings are:

- BitTorrent network traffic can have both seasonal and cyclical patterns.
- The ARMA time series can be used to provide short term forecasting of BitTorrent network traffic.
- The ARMA (2, 1) is better suited for cyclical BitTorrent network traffic patterns.
- The ARMA (3, 0) is well suited for seasonal BitTorrent network traffic patterns.

For our future works, we intend to investigate further on BitTorrent network traffic with Autoregressive Integrated Moving Average (ARIMA) as well as alternative routing methods to improve the overall BitTorrent network performance. We will focus on the unfairness issue that occurs in a P2P network where it reduces the overall performance of the network. This research is part of our comparative study on BitTorrent traffic pattern and we are confident that it can be appliedonBitTorrent client applications in the future.

**Authors**

**Poo KuanHoong** received his B.Sc. and M.IT. degrees in 1997 and 2001 respectively from the UniversitiKebangsaan Malaysia, Malaysia, and his Ph.D. degree in 2008 from the Department of Computer Science and Engineering, Nagoya Institute of Technology, Japan. Since June 2001, he has been a lecturer in the Faculty of Computing and Informatics at Multimedia University, Malaysia. His research is in the areas of peer-to-peer networks and distributed systems. He is a member of IEICE, IEEE and ACM

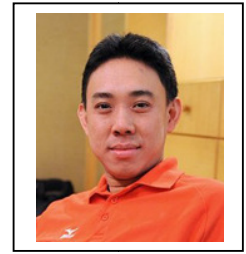

**Ian K. T. Tan** received his Master of Science in Parallel Computers and Computation from University of Warwick, United Kingdom and his Bachelor of Engineering in Information Systems Engineering from Imperial College London, United Kingdom in 1993 and 1992 respectively. He is currently a lecturer at Multimedia University, Cyberjaya; he has worked in the areas of microprocessor manufacturing, enterprise UNIX servers, storage management, software development to VoIP technologies. He is currently pursuing his Ph.D. at Faculty of Computing and Informatics, Multimedia University. His area of research is in operating systems scheduling for multicore processors, data transfer optimization across wide area networks and text processing. Mr Tan is a member is IACSIT, ACM, IEEE and an Associate of the City and Guilds Institute. He is Novell Certified Linux Administrator (NCLA), Novell Certified Linux Professional (NCLP).

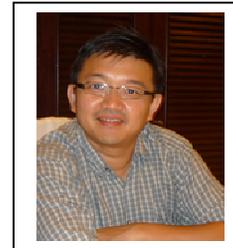

**CheeYikKeong** received his B.S. degree in Software Engineering from Multimedia University, Malaysia in 2010. He is a M.Sc. candidate in the Faculty of Computing and Informatics, Multimedia University, Malaysia. His current research interests include peer-to-peer & video streaming, distributed computing and networking

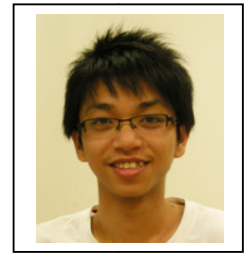